\documentclass[aps,preprint,nofootinbib]{revtex4-1}
\usepackage{graphicx}
\begin{document}
\title{\Large \bf Black String and G\"{o}del type Solutions of
 Chern-Simons Modified Gravity}
\author{\large Haji Ahmedov and Alikram N. Aliev}
\address{Feza G\"ursey Institute, \c Cengelk\" oy, 34684   Istanbul, Turkey}
\date{\today}

\begin{abstract}

Chern-Simons (CS) modified gravity with a prescribed CS scalar field does not admit rotating black hole solutions with spherical topology of the horizon. In this paper, we show that it does admit rotating {\it black hole/string} solutions with cylindrical topology of the horizon and present two  intriguing physical examples of such  configurations. First, we show that the Banados-Teitelboim-Zanelli (BTZ) stationary black string, that is obtained by adding on  a spacelike flat dimension to the BTZ black hole metric of three-dimensional gravity, solves the field equations of CS modified gravity with a specific source term and {\it irrespective of the choice of CS scalar field}. Next, we consider the Lemos solution for a rotating  straight black string in general relativity and show that for the CS scalar field being a function of the radial coordinate alone, this solution persists in CS modified gravity. We also discuss two examples of G\"{o}del type  metrics in CS modified gravity by uplifting to four dimensions a general one-parameter family of G\"{o}del type solutions of three-dimensional gravity. The first example is the usual  G\"{o}del  solution of general relativity which also survives in CS modified gravity with the CS scalar field depending on two variables, the radial and the azimuthal coordinates. The second example represents a new nontrivial (non general relativity)  G\"{o}del type solution to the vacuum field equations of CS modified gravity. This solution originates from the respective vacuum solution of topologically massive gravity when extending it to four dimensions by adding on an extra spatial coordinate and choosing the CS scalar field as a linear function of this coordinate.

\end{abstract}


\maketitle

\section{Introduction}

Chern-Simons modified gravity is an extension of general relativity due to a parity-violating correction term given by the Pontryagin density \cite{lue,jackiw}. After all, such a modification has its physical roots  in  string theory . As is known, the low-energy limit of string theory comprises  the Einstein-Hilbert action  supplemented  by the Pontryagin  term, which  is necessary  for canceling gravitational anomaly in the theory \cite{green} (for more details see a  review paper \cite{ay1} and references therein). On the other hand, the search for a consistent theory of quantum gravity has  stimulated studies of lower dimensional gravity models. Among these models, Deser, Jackiw and Templeton's theory of  topologically massive gravity (TMG) is of considerable interest \cite{djt}. In contrast to three-dimensional general relativity, TMG  possesses  a propagating degree of freedom  with a single massive graviton. Jackiw and Pi showed that one can also arrive at CS modified gravity within a geometrical framework, extending  a parity-violating gravitational Chern-Simons term of TMG into four dimensions \cite{jackiw}. This results in the gravitational action  which, in addition to the usual Einstein-Hilbert term, also involves the Pontryagin term coupled to a prescribed scalar field (a CS scalar field).  As a consequence, the field equations of CS modified gravity consist of two sectors: the Einstein  and  Chern-Simons sectors. In the latter case, the defining quantity is a second-rank, symmetric and traceless tensor that can be formally thought of as  a four-dimensional ``cousin" of  the  three-dimensional Cotton tensor. Following the authors of \cite{jackiw}, we shall call it the four-dimensional Cotton tensor.\footnote{This notion differs from that given in \cite{gh}} It is also important to  note that the field equations of CS modified  gravity imply an additional constraint equation (the vanishing of the Pontryagin density).

It is clear that the Pontryagin constraint would restrict the class of possible exact solutions to CS modified gravity. For instance, the Pontryagin density is not zero for the familiar Kerr solution
and therefore, CS modified gravity does not support this solution.
On the other hand, the Schwarzschild and  Reissner-Nordstr\"{om} metrics  fulfill the Pontryagin constraint and for a canonical choice of the CS scalar field (when it is a linear function of time alone) these metrics are solutions of CS modified gravity as well \cite{jackiw, guar}. In further developments,  some stationary metrics that could serve as an analogue of the Kerr solution in CS modified gravity have been  discussed in a number of works. For instance, in Refs. \cite{ay2,ay3} a stationary, but non-axisymmetric solution was  given in the far-field approximation and with the canonical choice of the CS scalar field. Meanwhile, a stationary and axisymmetric metric in the limit of slow rotation and for a special (noncanonical) choice of the CS scalar field was found in \cite{konno1}. It is worth noting that though these solutions are of importance to figure out some gravitomagnetic effects of CS modified gravity, none of them does correspond to the desired  analogue of the Kerr metric.

A systematic attempt to find  rotating  black hole solutions to CS modified gravity  was undertaken in \cite{grumiller}. It was shown that for noncanonical choices of the CS scalar field,  there exist two types of  solutions: the ultrarelativistically boosted limit of the Kerr solution (the Aichelburg-Sexl limit) and a stationary axisymmetric solution belonging to the van Stockum subclass of general stationary axisymmetric metrics.  Furthermore, the authors  of \cite{grumiller}
argued  against the existence of an exact counterpart of the Kerr solution, even for the most general prescribed CS scalar field. However, the situation may be different in the dynamical framework, where the scalar field evolves being driven by the Pontryagin density. Remarkably, such an analytical solution for rotating black holes was given in recent papers \cite{pretorius,konno2}. This solution is given in the slow-rotation limit of the Kerr metric (up to second order in rotation parameter) which also involves a small CS correction term of  the same order of magnitude. A similar problem was earlier studied in the context of some string-inspired models, where the Chern-Simons term serves as a source for axion fields in the background of the Kerr metric \cite{olive, reuter}. However, it should be noted that the exact Kerr solution with Chern-Simons  ``hairs"  still remains elusive.

In the following, we shall focus only on the theory CS modified gravity with a prescribed  CS scalar field \cite{jackiw}. Motivated by the fact that this theory does not admit rotating black hole solutions with  spherical topology of the horizon for instance, the Kerr solution, we ask the following natural question: {\it Does it support rotating black hole type solutions  with the horizon topology different from the spherical one}?  It is the main purpose of this paper to answer this question. As we have mentioned above, one can arrive at CS modified gravity  by embedding the Chern-Simons term of TMG (along with  an appropriate embedding coordinate: the gradient of the CS scalar field)
into four-dimensional spacetime. That is, in some sense,  CS modified gravity can be thought of as a four-dimensional ``counterpart" of TMG. As is known \cite{kaloper},  TMG  with a negative cosmological constant supports the  stationary BTZ black hole  solution \cite{btz1, btz2} which is thought to be an ``analogue" of the Kerr  solution in three-dimensional gravity. The physical importance  of the BTZ black holes is related to their indispensable  role, as ``theoretical laboratories", in understanding  the classical and quantum nature of gravity in three dimensions, especially in the context of the  AdS/CFT correspondence (see, for instance,  a recent paper \cite{strom} and references therein). We recall that this correspondence relates the properties of gravity in anti-de Sitter (AdS) background to those of dual conformal quantum field theory (CFT) residing on its boundary \cite{mkpw}.

With all this in mind, it is of great importance to look for BTZ type solutions, as  cylindrical black hole/string configurations, in CS modified gravity with a negative cosmological constant.  Of course, from the astrophysical point of view  such solutions describe an idealized situation, though within general relativity it has been argued that cylindrical collapse of an appropriate matter in the background  of the negative cosmological constant  will form the black string configurations \cite{lemos1, lemos2, lemos3}. However, such solutions would be certainly important in understanding the classical and quantum structures of CS modified gravity with the negative cosmological constant, at least, in the sense of  translating the unusual classical/quantum properties of their counterparts in three-dimensional gravity into four dimensions.

In this paper, we show that CS modified gravity does indeed admit this type of  rotating  black string configurations. First, we consider  the stationary BTZ black string,  that is obtained by adding on an extra spacelike flat dimension to the metric of the three-dimensional BTZ black hole, and show that it solves the field equations of CS modified gravity  with a specific source term determined by a negative cosmological constant and   regardless of the form of CS scalar field.  We note that  such a  cylindrical configuration was earlier considered in  \cite{lemos1} within general relativity. Next, we consider  a more general stationary  black string solution (or cylindrically symmetric rotating black hole solution)  found by Lemos  \cite{lemos2} in general relativity. In contrast to  the BTZ black string, the existence of  the Lemos black string configuration  in CS modified gravity depends on the form of CS scalar field. We show that for a CS scalar field depending on the radial coordinate alone,  the Lemos black string configuration is  supported by CS modified gravity as well.

Finally, we discuss two examples of G\"{o}del type  metrics in CS modified gravity by uplifting to four dimensions a general one-parameter family of G\"{o}del type solutions to three-dimensional gravity with a negative cosmological constant and with a matter source, corresponding to a uniform pressureless dust \cite{vuorio}. In the first example, the value of the parameter becomes fixed to $\lambda =\sqrt{2} $  when uplifting this solution to four dimensions through adding on a spacelike flat dimension. This results in the ordinary G\"{o}del solution \cite{godel} of general relativity. We show that the G\"{o}del solution survives in CS modified gravity with the CS scalar field depending on two variables, the radial and the azimuthal coordinates. Thus, in this context we  provide  a simple   derivation of the result of \cite{santos}. In the second example, we present a new nontrivial (non general relativity)  G\"{o}del type solution to the vacuum field equations of CS modified gravity. This solution is obtained by adding on a flat spatial coordinate to the three-dimensional G\"{o}del type metric with the parameter $ \lambda=2, $ that solves the vacuum field equations of TMG,  and choosing the CS scalar field as a linear function of this coordinate.

The paper is organized as follows: In Sec.II we present the field equations of CS modified gravity with a cosmological constant and with a matter term  and discuss their basic properties. In Sec.III we  study the BTZ and the Lemos black string configurations that solve the field equations of CS modified gravity with the negative cosmological constant. In Sec.IV  we discuss two examples of G\"{o}del type  metrics in CS modified gravity: The first is the usual  G\"{o}del  solution of general relativity and the second  is a new nontrivial  G\"{o}del type solution that exists in the absence of the cosmological constant.

\section{CS Modified Gravity}

The total action  for CS modified gravity with inclusion of a matter term is given by
\begin{eqnarray}
S &=& \frac{1}{16\pi G} \int d^4x \sqrt{-g}\left(R - 2 \Lambda + \frac{1}{4} \, \vartheta \,^\ast{R} R\right) + S_{mat}\,,
\label{csaction}
\end{eqnarray}
where the usual Hilbert term  with the cosmological constant $ \Lambda $ is supplemented with the Pontryagin density
\begin{eqnarray}
^\ast{R} R &=& {^\ast{R}^{\mu}_{~\,\nu}}^{\lambda \tau} R^{\nu}_{~\,\mu \lambda \tau}\,,
\label{pontryagin}
\end{eqnarray}
coupled to the CS scalar field  $ \vartheta  $  and $ S_{mat} $  is the matter action. The dual tensor in (\ref{pontryagin}) is defined as
\begin{eqnarray}
{^\ast{R}^{\mu}_{~\,\nu}}^{\lambda \tau} &=& \frac{1}{2} \, \epsilon^{\lambda \tau \rho \sigma } R^{\mu}_{~\,\nu \rho \sigma}\,,
\label{dual}
\end{eqnarray}
where the Levi-Civita tensor is given by $ \epsilon^{\,\lambda \tau \rho \sigma}= \varepsilon^{\,\lambda \tau \rho \sigma}/\sqrt{-g},~~\varepsilon^{0123}=1  $.

It is curious that the Pontryagin  density can be written as  the total divergence of  a four-dimensional  topological current. We have
\begin{eqnarray}
^\ast{R} R &= & 2\,\nabla_{\mu}K^{\mu}\,,
\label{divcurrent}
\end{eqnarray}
where
\begin{eqnarray}
K^{\mu} &= & \epsilon^{\mu \nu \alpha \beta} \Gamma^{\lambda}_{\nu \tau}\left(\partial_{\alpha}\Gamma^{\tau}_{\beta \lambda}
+\frac{2}{3}\,\Gamma^{\tau}_{\alpha \sigma}\Gamma^{\sigma}_{\beta \lambda}\right)\,
\label{current}
\end{eqnarray}
and  the quantities $ \Gamma^{\mu}_{\alpha \beta} $  are the Christoffel symbols.

Varying action (\ref{csaction}) with respect to  the spacetime metric, we obtain the field equations
\begin{eqnarray}
G_{\mu\nu} + \Lambda g_{\mu\nu} + C_{\mu\nu} &= & 8\pi T_{\mu\nu}\,,
\label{fieldeqs}
\end{eqnarray}
where the Einstein tensor
\begin{eqnarray}
G_{\mu\nu} &= & R_{\mu\nu} - \frac{1}{2}\, R g_{\mu\nu}\,,
\label{Etensor}
\end{eqnarray}
$ T_{\mu\nu} $ is the energy-momentum tensor of matter and $ C_{\mu\nu} $ the four-dimensional ``Cotton" tensor which is symmetric and traceless \cite{jackiw}. The contravariant  Cotton tensor is given by
\begin{eqnarray}
C^{\mu\nu} &= & \vartheta_{\alpha}\,\epsilon^{\,\alpha \beta \gamma(\mu} R^{ \nu)}_{ \beta \,;\gamma}
+\vartheta_{\alpha \beta}\,^\ast{R}^{\beta (\mu \nu)\alpha}\,\,,
\label{cotton4d}
\end{eqnarray}
where the semicolon denotes covariant differentiation and
\begin{eqnarray}
\vartheta_{\alpha}& = & \vartheta_{;\alpha} ~~~~~ \vartheta_{\alpha \beta}=  \vartheta_{\alpha ;\beta}\,.
\label{embcoord}
\end{eqnarray}
Clearly, one can also perform the variational procedure in (\ref{csaction}) with respect to the CS scalar field (thinking of it as a dynamical variable). This results in the constraint  equation
\begin{eqnarray}
 ^\ast{R} R &= & 0\,.
\label{pconstraint}
\end{eqnarray}
Alternatively, taking the divergence of equation (\ref{fieldeqs}) and comparing the result with the fact that
\begin{eqnarray}
\nabla_{\nu} C^{\mu\nu} &= & - \frac{1}{8}\,\vartheta^{\mu}\, ^\ast{R} R\,,
\label{cotdiv}
\end{eqnarray}
one can see that the contracted Bianchi identities along with the energy-momentum conservation lead to equation (\ref{pconstraint}) as well. Thus,  the field equations of CS modified gravity (\ref{fieldeqs}) must be accompanied  by  the Pontryagin constraint (\ref{pconstraint}). Clearly, the Pontryagin constraint will be fulfilled only for a certain class of spacetimes. As we have already mentioned in the introductory section, it is automatically satisfied for the Schwarzschild and  Reissner-Nordstr\"{om} metrics which are  solutions to CS modified gravity for the most general choice of the CS scalar field (see \cite{grumiller} for details). Comparison of equations (\ref{pconstraint}) and (\ref{cotdiv}) shows that equation (\ref{pconstraint}) gives only a necessary  condition  for a spacetime to be a solution of CS modified gravity.

\section{Rotating black strings in CS modified gravity}

In this section, we  discuss two examples of  stationary  black string configurations  that solve the field equations of CS modified gravity with a negative cosmological constant. The first example is the BTZ black string configuration, which is obtained by adding on a spacelike flat dimension to the metric of a three-dimensional BTZ black hole. Another example is the stationary black string (or cylindrical black hole) solution of general relativity  found by Lemos in \cite{lemos2}.

\subsection{The BTZ black string}

We begin by recalling some properties of the BTZ black hole \cite{btz1,btz2}. First of all, this is an anti-de Sitter type   solution of three-dimensional general relativity. Though the curvature is locally constant for this solution, the black hole structure arises when identifying inequivalent points of the anti-de Sitter  space under a discrete subgroup of its  isometry group $ SO(2,2) $. As a consequence, the quotient space represents a  black hole with two physical parameters:  the mass  $ M $ and  the angular momentum $ J $. The metric is given by
\begin{eqnarray}
^{(3)}ds^2 & = &  \left(M-\frac{r^2}{l^2}\right) dt^2 + \left(\frac{r^2}{l^2}- M +\frac{J^2}{4r^2}\right)^{-1} dr^2+r^2 d\phi^2- J dt d\phi\,,
\label{btzmetric}
\end{eqnarray}
where we have introduced the length parameter $ l^{-2}= -\Lambda $. This is a crucial length parameter for the  horizon formation in a three-dimensional theory, where the mass is a dimensionless.  For $ M=-1 $ and  $ J=0 $ we have  the usual anti-de Sitter metric. The horizon stricture of ({\ref{btzmetric}) is governed by the equation
\begin{eqnarray}
\frac{r^2}{l^2}- M +\frac{J^2}{4r^2}&=& 0\,,
\label{horeq}
\end{eqnarray}
which has two roots  $ r_ {+} $ and $ r_ {-} $ corresponding to the radii of outer and inner horizons, respectively. We have
\begin{eqnarray}
r^2_{\pm}&=& \frac{M l^2}{2}\left(1\pm \sqrt{1- \frac{J^2}{M^2l^2}}\,\right)\,.
\label{horradi}
\end{eqnarray}
Cosmic censorship requires that
\begin{eqnarray}
|J|\leq M l\,,
\label{cs}
\end{eqnarray}
where the equality defines the extremal horizon of the
black hole.

It is interesting that the BTZ metric (\ref{btzmetric}) is  also a trivial solution to TMG  as the three-dimensional Cotton tensor vanishes identically for this metric \cite{kaloper}. Furthermore, the BTZ black string configuration, as an extension of metric (\ref{btzmetric}) to four dimensions by adding on an extra flat direction,  is also known in ordinary general relativity \cite{lemos1}. With this in mind, and taking into account the fact that CS modified gravity can be obtained by uplifting (with an appropriate embedding coordinate)  TMG to four dimensions \cite{jackiw} it is natural to ask: {\it Does the BTZ black string configuration persist in CS modified  gravity}?   In the following, we answer this question, showing that the BTZ black string configuration described by the metric
\begin{eqnarray}
ds^2 & = &  h_{\mu\nu} dx^{\mu} dx^{\nu} + dz^2
\label{btstring}
\end{eqnarray}
is also a solution to CS modified  gravity with an appropriate source term and regardless of the form of CS scalar field. Here  $ h_{\mu\nu} $ denotes the three-dimensional metric given in  (\ref{btzmetric}).

It is straightforward to show that  metric (\ref{btstring}) fulfils the Pontryagin constraint (\ref{pconstraint}). Next, what we need is to show that the four-dimensional Cotton tensor (\ref{cotton4d}) vanishes identically for this metric. For this purpose, we first note that the cylindrical symmetry of this metric implies the existence of the Killing vector
\begin{equation}
\xi=\partial/\partial z\,,
 \label{killz}
\end{equation}
of constant length ($ \xi^2=1 $). Unlike the timelike Killing vector $ \partial/\partial t $, this vector  is hypersurface orthogonal, obeying the equation
\begin{equation}
\xi_{[\mu ; \nu} \,\xi_{\lambda]} = 0\,.
\label{killingh}
\end{equation}
Thus, we can use the results of a recent work \cite{ah} to  examine  all possible projections of the Cotton tensor (\ref{cotton4d}) in directions parallel and orthogonal to the Killing vector (\ref{killz}). In \cite{ah}, it was shown that the projection of this tensor in direction parallel to a hypersurface orthogonal Killing vector vanishes identically.
That is,
\begin{eqnarray}
C_{**} & = & C_{\mu \nu} \xi^{\mu} \xi^{\nu}  \equiv  0\,.
\label{paralcot1}
\end{eqnarray}
The mixed  projection of the Cotton tensor is given by
\begin{eqnarray}
\overline{C}_{*\lambda}&=&  C_{\mu \nu}\xi^{\mu}h^{\nu}_{\;\lambda}\,,
\label{parpercot}
\end{eqnarray}
where $ h_{\;\nu}^{\mu} $ is the projection operator defined as
\begin{eqnarray}
h_{\;\nu}^{\mu} &=& \delta_{\;\nu}^{\mu}
    - \xi^{\mu} \xi_{\nu}  \,,~~~h_{\;\nu}^{\mu} \,\xi^\nu = 0 \,,~~~
h_{\;\lambda}^{\mu} \, h_{\;\nu}^{\lambda}  = h_{\;\nu}^{\mu}\,.
\label{proj1}
\end{eqnarray}
Using now equation (\ref{cotton4d}) in (\ref{parpercot}) and taking into account the equations
\begin{eqnarray}
^{(3)}R_{\mu \nu}&=& -\frac{2}{l^2} \,h_{\mu \nu}\,,~~~~R_{\mu \nu}= -\frac{2}{l^2} \left(g_{\mu \nu}-\xi_{\mu}\xi_{\nu}\right),
\label{ric3}
\end{eqnarray}
for metrics (\ref{btzmetric}) and (\ref{btstring}), respectively, we find that
\begin{eqnarray}
\overline{C}_{*\lambda}& =& \frac{1}{2}\,\vartheta_{\alpha \beta} \xi_{\mu} \epsilon^{\,\mu \alpha \rho \tau}  {R^{\beta}}_{\nu\rho\tau}  h_{\;\lambda}^{\nu}\,.
\label{parpercot1}
\end{eqnarray}
In obtaining this expression we have also used the fact
\begin{equation}
\xi_{\mu ;\nu} =0\,,~~~~~\xi_{\tau}  R^{\tau}\,_{\lambda \nu \mu}=0\,
\label{cdkilling}
\end{equation}
for the hypersurface orthogonal Killing  vector of constant length (see Ref.\cite{ah} for details). With this in mind and using the equation
\begin{eqnarray}
^{(3)}R_{\mu \nu \lambda \tau} & = & h^{\alpha}_{\;\mu}h^{\beta}_{\;\nu} h^{\rho}_{\;\lambda} h^{\sigma}_{\;\tau} R_{\alpha\beta\rho\sigma}
=-\frac{1}{l^2}\left(h_{\mu\lambda}h_{\nu \tau}- h_{\mu\tau}h_{\nu \lambda}\right),
\label{cur3to4}
\end{eqnarray}
one can further transform  expression (\ref{parpercot1}) into the form
\begin{eqnarray}
\overline{C}_{*\lambda}& =& -\frac{1}{2l^2}\,\vartheta_{\alpha}^{\,\,\, \beta} \xi_{\mu} \epsilon^{\,\mu \alpha \rho \tau} \left(h_{\beta\rho}h_{\lambda \tau}- h_{\beta\tau}h_{\lambda\rho}\right).
\label{parpercotf}
\end{eqnarray}
Taking in this expression into account equation (\ref{proj1})  it is easy to see that
\begin{eqnarray}
\overline{C}_{*\lambda}  \equiv  0\,.
\label{mixedzero}
\end{eqnarray}
The remaining step is to show that the orthogonal projection of the Cotton tensor (\ref{cotton4d}) vanishes identically as well. We appeal to a general expression for this projection given in \cite{ah},  which in the case under consideration, can be written in the form
\begin{eqnarray}
\overline{C}^{\,\lambda \sigma} & = & \xi_{\tau}\,\left[
(\vartheta_{\alpha}\xi^{\alpha})\,\epsilon^{\,\tau \beta \gamma (\lambda}D_{\gamma}\overline{R}^{\,\sigma)}_{\beta}- \vartheta_{\alpha \beta}\xi^{\alpha}
\epsilon^{\,\tau \beta \gamma (\lambda}\overline{R}^{\,\sigma)}_{\gamma} \right],
\label{finalcotperp}
\end{eqnarray}
where $D_{\mu} $ is the derivative operator with  respect to the three-dimensional metric and
\begin{eqnarray}
\overline{R}_{\mu\nu} & = & h^{\alpha}_{\;\mu}h^{\beta}_{\;\nu} R_{\alpha\beta} =  \,^{(3)}R_{\mu \nu}\,.
\label{ricci3to4}
\end{eqnarray}
Comparing these equations with the first equation in (\ref{ric3}), we immediately see that
\begin{eqnarray}
\overline{C}^{\,\lambda \sigma} \equiv  0\,.
\label{finalzero}
\end{eqnarray}
Thus, we obtain that  all components of the Cotton tensor vanish identically for the BTZ black string configuration (\ref{btstring}). This confirms that the BTZ black string is a  solution of the field equation (\ref{fieldeqs}) with the specific source term
\begin{equation}
8\pi T_{\mu\nu} = \frac{2}{l^2}\,\xi_{\mu}\xi_{\nu}\,,
\label{spsourcebtz}
\end{equation}
and irrespective of the form of CS scalar field.

\subsection{The Lemos black string}

It is a remarkable fact that the BTZ black string  is not the only example that relates three-dimensional gravity with a negative cosmological constant to four-dimensional general relativity with cylindrical symmetry. Lemos \cite{lemos2} showed that there exists a ``generic" stationary black string solution to general relativity with  the negative cosmological constant. When reduced to three dimensions through dimensional reduction procedure, this solution corresponds to a black hole in three-dimensional dilaton-gravity  (see also \cite{horowitz}). The metric of the Lemos black string can be written in the form
\begin{eqnarray}
ds^2 & = &-\left(\frac{r^2}{l^2}-\frac{M}{r} \,\Xi^2\right)dt^2 +\left(\frac{r^2}{l^2}-\frac{M}{r}\right)^{-1}dr^2 +\left(r^2+ \frac{M}{r}\,a^2\right)d\phi^2 \nonumber \\[2mm]  & &
-\frac{2M a}{r} \,\Xi \,d\phi\, dt +\frac{r^2}{l^2}\,dz^2\,,
\label{lemosstring}
\end{eqnarray}
where  the  length parameter $ l^{2}= -3\Lambda^{-1} $ and
\begin{equation}
\Xi = \left(1+\frac{a^2}{l^2}\right)^{1/2}.
\label{spsource}
\end{equation}
The parameters $ M $ and   $ a $  can be related to the physical mass and angular momentum, respectively. It is important to note that the latter quantities, just as in the case of BTZ string, must be defined per unit length of the string. In \cite{lemos2, lemos3}, the mass and angular momentum  were consistently defined by passing to an equivalent three-dimensional theory and employing the well-known Hamiltonian approach with Brown-York prescription \cite{by}. The horizon and ergosphere structures are given by equations $ g^{rr}=0 $ and $ g_{tt}=0 $, respectively.

A simple coordinate transformation of the form
\begin{eqnarray}
T= \Xi \,t - a \phi\,,~~~~~ \varphi = \frac{a}{l^2}- \Xi \phi\,,
\label{trans}
\end{eqnarray}
enables one to reduce  metric (\ref{lemosstring})  to that of a static spacetime.  However, in general such an identification is not valid as the spacetime is not simply connected (the first Betti number is one) and there are no global transformations mapping static and stationary black string spacetimes. A detailed  analysis of the global structure of metric (\ref{lemosstring}) can be found in \cite{lemos3}.

It is easy to see that metric (\ref{lemosstring}) possesses
time-translational  and  rotational Killing vectors
\begin{equation}
\xi_{(t)}=\frac{\partial}{\partial t}\,\,, ~~~~~~~~~\xi_{(\phi)}=
\frac{\partial}{\partial \phi}\,\,,
\label{skillings}
\end{equation}
as well as the Killing vector given in (\ref{killz})  that corresponds to the translational symmetry along $z$-axis. This vector is hypersurface orthogonal. However, in the case under consideration, its length $ \xi^2 \neq const $.

In the following, we show that for some choice of the CS scalar field the black string solution (\ref{lemosstring}) solves equations of CS modified gravity as well.  First of all, it is easy to check that this solution satisfies the Pontryagin constraint (\ref{pconstraint}). Next, we use the decoupling theorem of \cite{ah} which  states that  for a hypersurface orthogonal Killing vector and for a CS scalar field being constant along this vector, the source-free  equations of CS modified gravity  decouple into their Einstein and Cotton sectors. Thus, this theorem leads to the following  independent equations
\begin{eqnarray}
E_{\mu\nu} &= & G_{\mu\nu} -\frac{3}{l^2}\, g_{\mu\nu}=0\,, ~~~~~~~  C_{\mu\nu} = 0\,,
\label{fielddec}
\end{eqnarray}
with the CS scalar field, obeying the condition
\begin{eqnarray}
\pounds_{\xi} \vartheta  & = &  \left(\vartheta_{\alpha}\xi^{\alpha}\right) = 0\,,
\label{orthscalar}
\end{eqnarray}
where $ \pounds_{\xi} $ is the Lie derivative along the hypersurface orthogonal  Killing vector $ \xi  $.  In fact,  exploring  for each equation in (\ref{fielddec}) all respective projections   in directions along and orthogonal to the Killing vector \cite{ah}, we find that
\begin{eqnarray}
C_{**} & = & C_{\mu \nu} \xi^{\mu} \xi^{\nu}  \equiv  0\,,~~~ \overline{C}^{\lambda \sigma} =  C^{\mu\nu} h^{\lambda}_{\;\mu} h^{\sigma}_{\nu} \equiv  0\,, ~~~\overline{E}_{*\lambda} =  E_{\mu \nu}\xi^{\mu}h^{\nu}_{\;\lambda} \equiv  0\,.
\label{ids1}
\end{eqnarray}
With these identities, the remaining projections vanish as a consequence of the field equations. Thus, we have
\begin{eqnarray}
E_{**} & = & E_{\mu \nu} \xi^{\mu} \xi^{\nu} =  0\,,~~~\overline{E}^{\lambda \sigma} =  E^{\mu\nu} h^{\lambda}_{\;\mu} h^{\sigma}_{\nu} = 0\,.
\label{eqs1}
\end{eqnarray}
and
\begin{eqnarray}
\overline{C}_{*\lambda}&=&  C_{\mu \nu}\xi^{\mu}h^{\nu}_{\;\lambda} =0\,.
\label{eqs2}
\end{eqnarray}

It is straightforward to check that  metric (\ref{lemosstring}) satisfies equations in (\ref{eqs1}) as required by its very existence.
Meanwhile, writing out for this metric all components of equation (\ref{eqs2}) explicitly, we obtain that they are equivalent to vanishing of the Cotton tensor components
\begin{eqnarray}
C^{t z}& = & - \frac{3Ml}{2r^6}\,\Xi \left(\Xi\, \partial_{\phi} + a \partial_{t}\right)\left(r \partial_{r}\vartheta - \vartheta\right),\\[2mm]
C^{r z}& = & \frac{3Ml}{2r^5}\, \left(\Xi \,\partial_{t} + \frac{a}{l^2} \, \partial_{\phi}\right)\left(\Xi\, \partial_{\phi}\vartheta + a \partial_{t} \vartheta\right),\\[2mm]
C^{\phi z} & = & - \frac{3Ma}{2r^6 l}\, \left(a \,\partial_{t} + \Xi \, \partial_{\phi}\right)\left(r \partial_{r}\vartheta - \vartheta\right),
\label{nonvanishcots}
\end{eqnarray}
which allow to specify the form of the CS scalar field as
\begin{eqnarray}
\vartheta &= & \vartheta(r)\,.
\label{CSfield1}
\end{eqnarray}
Thus, with this CS scalar field the black string solution  of general relativity  given in (\ref{lemosstring}) is also a solution to CS modified gravity.

\section{G\"{o}del type solutions}

We now examine G\"{o}del type solutions to CS modified gravity by uplifting  a one-parameter family of three-dimensional G\"{o}del type metrics \cite{vuorio} to four dimensions. We begin with the metric
\begin{eqnarray}
^{(3)}ds^2=-\left[dt-\lambda\left(\cosh r -1\right)d\phi\right]^2 + dr^2 + \sinh^2r \,d\phi^2\,,
\label{3dvugo}
\end{eqnarray}
where $\lambda $ is an arbitrary constant parameter. Calculating  for this metric the nonvanishing mixed  components of the Ricci tensor, we find that
\begin{eqnarray}
^{(3)}R^{t}_{\,\,t}&= &-\frac{\lambda^2}{2}\,,~~~~
^{(3)}R^{r}_{\,\,r}= \, ^{(3)}R^{\phi}_{\,\,\phi}=-1+\frac{\lambda^2}{2}\,,~~~~ ^{(3)}R^{t}_{\,\,\,\phi}= \lambda \left(\lambda^2-1\right) \left(\cosh r -1 \right).
\label{vugosricgen}
\end{eqnarray}
It follows that the Ricci scalar
\begin{eqnarray}
^{(3)} R &=& -2+\frac{\lambda^2}{2}\,\,.
\label{vugosc}
\end{eqnarray}
With this quantities it is not difficult to verify that metric (\ref{3dvugo}) satisfies the three-dimensional Einstein field equations \begin{eqnarray}
^{(3)}G^{\mu}_{\,\,\, \nu}+ \Lambda \,^{(3)}\delta^{\mu}_{\,\,\, \nu}  &= & 8\pi \,^{(3)}T^{\mu}_{\,\,\, \nu}\,,
\label{3deinscos}
\end{eqnarray}
with the source term
\begin{eqnarray}
^{(3)}T^{\mu\nu}= \frac{\lambda^2-1}{8\pi} \,v^{\mu} v^{\nu}\,,~~~~~v^{\mu}= (1,0,0)\,,
\label{lamrho}
\end{eqnarray}
corresponding to a uniform pressureless dust matter and with the negative cosmological constant
\begin{eqnarray}
\Lambda &= & -\frac{\lambda^2}{4}\,.
\label{lamcosm1}
\end{eqnarray}
It has been shown that that metric (\ref{3dvugo}) possesses all
peculiar properties of original four-dimensional G\"{o}del spacetime \cite{godel}. Thus, for any nonzero value of $\lambda $, this metric represents G\"{o}del  type solutions in three dimensions \cite{vuorio}. The discussion of  more general three-dimensional G\"{o}del  type  solutions  can be found in \cite{gurses}.

Next, we extend metric (\ref{3dvugo}) to four dimensions by adding on an extra spacelike flat dimension. That is, we consider the metric in the form
\begin{eqnarray}
ds^2=-\left[dt-\lambda\left(\cosh r -1\right)d\phi\right]^2 + dr^2 + \sinh^2r \,d\phi^2 + dz^2\,.
\label{4vugo}
\end{eqnarray}
This metric, just like  those in the previous cases (see Sec.III),
satisfies the Pontryagin constraint (\ref{pconstraint}). Furthermore, it also admits the hypersurface orthogonal Killing vector (\ref{killz}) of length $\xi^2=1 $ . This means that with the CS scalar field given in (\ref{orthscalar}), the field equations of CS modified gravity (\ref{fieldeqs}) decouple into the Einstein and Cotton sectors, according to the decoupling theorem of \cite{ah}. Thus, we have two independent equations
\begin{eqnarray}
G_{\mu\nu} + \Lambda g_{\mu\nu}= 8\pi T_{\mu\nu}\,, ~~~~~~~  C_{\mu\nu} = 0\,,
\label{fieldgodcs}
\end{eqnarray}
where
\begin{eqnarray}
\overline{T}_{* \lambda}= T_{\mu\nu} \xi^{\mu} h_{\;\lambda}^{\nu} \equiv  0 .
\label{mixed}
\end{eqnarray}
Clearly,  the four-dimensional counterpart of (\ref{lamrho})  given by
\begin{eqnarray}
T^{\mu\nu}= \rho \,u^{\mu} u^{\nu}\,,~~~~u^{\mu}= (1,0,0,0)\,,
\label{4lamrho}
\end{eqnarray}
fulfils equation (\ref{mixed})  and
\begin{eqnarray}
T_{**} & = & T_{\mu \nu} \xi^{\mu} \xi^{\nu} =0\,.
\label{emtpdust}
\end{eqnarray}
With this in mind and using the fact that for any hypersurface Killing vector of constant length
\begin{eqnarray}
R_{**} & = & R_{\mu \nu} \xi^{\mu} \xi^{\nu} =0
\label{paralric1}
\end{eqnarray}
and $ ^{(3)}R= R \,$ (see Ref.\cite{ah}),  from the first equation in (\ref{fieldgodcs}),  we find that
\begin{eqnarray}
\Lambda=-1+\frac{\lambda^2}{4}\,.
\label{lamcosm2}
\end{eqnarray}
Comparison of this expression with that in (\ref{lamcosm1}) gives $\lambda =\sqrt{2}$. With this value of the parameter $ \lambda $, it is easy to show that metric (\ref{4vugo}) satisfies the Einstein field equations in (\ref{fieldgodcs}) with
\begin{eqnarray}
8\pi\rho & = & 1\,,~~~~\Lambda = - 4\pi\rho\,.
\label{gotmu}
\end{eqnarray}
On the other hand, computing the curvature invariants of metric (\ref{4vugo}) for $ \lambda =\sqrt{2}\,, $ we obtain
\begin{eqnarray}
R_{\mu\nu}R^{\mu\nu} & = & 1 \,,~~~~~ R_{\mu \nu \sigma \tau}R^{\mu \nu \sigma \tau}=3\,.
\label{4vugoinv}
\end{eqnarray}
These quantities  turn out to be precisely the same (up to overall constant factor) as those for the  G\"{o}del metric \cite{godel} in general relativity. It follows that a one-parameter family of G\"{o}del type three-dimensional spacetimes (\ref{3dvugo}), when extended to four dimensions as given in (\ref{4vugo}),  goes over into the only one spacetime with fixed value of $ \lambda =\sqrt{2} $  which, in essence, is nothing but the usual G\"{o}del solution \cite{godel}.

We  now examine the second equation in (\ref{fieldgodcs}). We recall that the identities  in (\ref{ids1}) hold in our case as well. That is, we have  $ C_{**} \equiv  0 $ regardless of the CS scalar field and  $ \overline{C}^{\lambda \sigma} \equiv  0 \,$  due to equation (\ref{orthscalar}). The remaining equation (\ref {eqs2}) is equivalent to the vanishing of the components
\begin{eqnarray}
C^{t z}& = & \frac{1}{\sqrt{2} \sinh r} \left[\left(\cosh r -1 \right)\partial_{t}\partial_{r}\vartheta + \sinh r \,\partial_{t} \vartheta \right],\\[2mm]
C^{r z}& = & -\frac{1}{2 \sinh r}\left[\sqrt{2} \left(\cosh r -1 \right) \partial_{t}^2 \vartheta  + \partial_{t} \partial_{\phi}\vartheta\right],\\[2mm]
C^{\phi z} & = & \frac{1}{2\sinh r}\,\partial_{t}\partial_{r}\vartheta\,.
\label{cotsvugo1}
\end{eqnarray}
It follows that the CS scalar field in the form
\begin{eqnarray}
\vartheta &= & \vartheta(r, \phi)\,
\label{CSfield2}
\end{eqnarray}
completes  vanishing of all components of the Cotton tensor. Thus, we see  that for the most general choice of the CS scalar field given in (\ref{CSfield2}), metric (\ref{4vugo}) with $ \lambda =\sqrt{2} $,  as a general relativity G\"{o}del solution of \cite{godel}, survives in CS modified gravity as well. This result agrees with that given in \cite{santos}. However, our description here is greatly simple due to the decoupling theorem of \cite{ah}.

To conclude this section, we also consider the value of $ \lambda =2 $ in metric (\ref{3dvugo}). In \cite{vuorio}, it was shown that such a metric (in units, in which $ m=3 $ ) satisfies the vacuum field equations of topologically massive gravity \cite{djt}
\begin{eqnarray}
^{(3)}R_{\mu \nu } \,+ \frac{1}{m} \, ^{(3)} C_{\mu\nu} &= & 0\,.
\label{tmgeqs}
\end{eqnarray}
The three-dimensional Cotton tensor is given by
\begin{eqnarray}
^{(3)}C^{\,\lambda \sigma} & = &
\epsilon^{\, \beta \gamma (\lambda}D_{\gamma} ^{(3)}{R}^{\,\sigma)}_{\beta}\,,
\label{cot3d}
\end{eqnarray}
where $  \epsilon^{\, \beta \gamma \lambda} $ is the three-dimensional Levi-Civita tensor. Thus, for $ \lambda =2 $ metric (\ref{3dvugo}) represents a G\"{o}del type solution not only  to the three-dimensional Einstein field equations (\ref{3deinscos}), but also to the vacuum field equations (\ref{tmgeqs}) of TMG.

In  our previous paper \cite{ah}, in addition to a decoupling theorem in CS modified gravity, we have also proved a reduction theorem which states: If a four-dimensional spacetime admits a non-null hypersurface orthogonal Killing vector of constant length and the gradient of the CS scalar field is parallel to the Killing vector, then CS modified gravity reduces to TMG in three dimensions. From this theorem it follows that having extended to four dimensions, the TMG solution for $ \lambda =2 $ would also solve the field equations of CS modified gravity in vacuum for some choice of the CS scalar field. Choosing, in accordance with this theorem,  the CS scalar as
\begin{eqnarray}
\vartheta &= & \alpha z\,,
\label{CSfieldlinear}
\end{eqnarray}
where $ \alpha $ is an arbitrary constant and calculating, for metric (\ref{4vugo}) with $ \lambda =2 $, the nonvanishing mixed components of the Ricci tensor, we find that
\begin{eqnarray}
-\frac{1}{2}\,R^{t}_{\,\,t}&= & R^{r}_{\,\,r}=R^{\phi}_{\,\,\phi}=1\,,~~~~~ R^{t}_{\,\,\phi}= 6 \left(\cosh r -1 \right)\,.
\label{ricnongrgod}
\end{eqnarray}
Similarly, for the nonvanishing components of the Cotton tensor in (\ref{cotton4d}) we have
\begin{eqnarray}
-\frac{1}{2}\,C^{t}_{\,\,t} & = & C^{r}_{\,\,r}=C^{\phi}_{\,\,\phi}=-3\alpha\,,~~~~~ C^{t}_{\,\,\phi}= -18 \alpha \left(\cosh r -1 \right)\,.
\label{cotnongrgod}
\end{eqnarray}
It is easy to see that for $ \alpha= 1/3 $ these quantities solve the field equations of CS modified gravity (\ref{fieldeqs}) in the vacuum case. That is, for  $ \lambda =2  $ the metric in (\ref{4vugo}) represents a  nontrivial (non general relativity) G\"{o}del type solution to the vacuum field equations of CS modified gravity.

\section{Conclusion}

CS modified gravity with a prescribed CS scalar field  does not support rotating black hole solutions with  spherical topology of the horizon. This occurs due to the restriction given by the Pontryagin constraint of the theory. For instance, the Kerr solution  of general relativity does not satisfy this constraint and therefore it is not a solution to CS modified gravity. In this paper, we have shown that the situation is different for rotating black holes with cylindrical topology of the horizon. Such black objects exist only in the presence of a negative cosmological constant and they are known as black strings. We have examined two physical examples of the stationary black string configurations.

We began with the BTZ black string configuration, known as a cylindrical system in four dimensions, which solves the Einstein field equations with a specific source term determined by the negative cosmological constant.  We have shown that for this configuration the  Cotton tensor vanishes identically and regardless of the form of CS scalar field. Thus, the BTZ black string  solves the field equation of CS modified gravity with the same specific source term as that in ordinary general relativity. Next, we have considered  the Lemos solution for a stationary black string in general relativity which on reducing to three dimensions, through dimensional reduction procedure, transforms into a black hole in three-dimensional dilaton-gravity \cite{lemos2}. We have  found that the Lemos black string  is also a solution to CS modified gravity with the CS scalar field chosen  as a function of the radial coordinate alone.

We have also considered two examples of  G\"{o}del type metrics in CS modified gravity by extending  a one-parameter family of G\"{o}del type solutions of three-dimensional gravity to four dimensions. Such a procedure uniquely specifies the value of the parameter to $ \lambda=\sqrt{2} $ and results in the usual G\"{o}del solution of general relativity. Within our description we have obtained  an earlier result of \cite{santos}, showing that for the CS scalar field being a function of two variables, the radial and the azimuthal coordinates, the G\"{o}del solution also satisfies the field equations of CS modified gravity. Meanwhile, fixing in the three-dimensional metric the value of the parameter to $ \lambda= 2 $ gives rise to  a  G\"{o}del type solution \cite{vuorio} to the vacuum field equations of TMG. We have shown that by adding on to this solution an extra  spatial flat coordinate and choosing the CS scalar field as a linear function of this coordinate, one arrives at a new nontrivial G\"{o}del type vacuum solution of CS modified gravity. Thus, CS modified gravity admits a non general relativity G\"{o}del type  solution with a characteristic  closed timelike curves (CTC's) structure, unlike the
G\"{o}del universe in string theory that needs not contain CTC's \cite{barrow}.

It should be emphasized that the stationary  black string solutions discussed above  are of importance for several reasons:  (i) First of all, they are the first physical examples of rotating exact metrics in  CS modified gravity with a prescribed CS scalar field. It turns out that the theory still supports black objects with rotational dynamics, though with the cylindrical horizon topology and in the presence of the negative cosmological constant, (ii) in the framework  of AdS/CFT correspondence  these solutions, just as their counterparts in three-dimensional gravity, may play an important role in understanding classical/quantum structure of CS modified gravity, (iii) in the astrophysical aspect,  the black string configurations may arise as endpoints of cylindrical collapse of an appropriate matter in the background  of the negative cosmological constant \cite{lemos1, lemos2, lemos3}. However, it is most likely that  they will describe an idealized astrophysical situation, (iv) though the black strings are general relativity  solutions supported by CS modified gravity as well, their physical content will be different, as they thermodynamical characteristics such as the mass, angular momentum and the entropy, just like in the case of the BTZ black holes in TMG \cite{park},
will be modified by the Chern-Simons contributions.  All together, these issues are challenging  tasks for future works.


\begin{thebibliography}{99}

\bibitem{lue}  A. Lue, L. M. Wang and  M. Kamionkowski, Phys. Rev. Lett. {\bf 83}, 1506 (1999).
\bibitem{jackiw} R. Jackiw and S. Y. Pi,  Phys. Rev. D {\bf 68}, 104012 (2003).
\bibitem{green} M. B. Green and J. H. Schwarz, Phys. Lett B {\bf 149}, 117 (1984).
\bibitem{ay1} S. Alexander and N. Yunes, Phys. Rept. {\bf 480}, 1 (2009).
\bibitem{djt} S. Deser, R. Jackiw and S. Templeton,
 Phys. Rev. Lett. {\bf 48}, 975 (1982); Ann. Phys. (N.Y.)
{\bf 140}, 372 (1982); {\it erratum-ibid}. {\bf 185} 406 (1988).
\bibitem{gh} A. Garcia, F. W, Hehl, C. Heinicke and A. Macias, Class. Quant. Grav. {\bf 21}, 1099 (2004).
\bibitem{guar} D. Guarrera and A. J. Hariton, Phys. Rev. D {\bf 76}, 044011 (2007).
\bibitem{ay2} S. Alexander and N. Yunes, Phys. Rev. D {\bf 75}, 124022 (2007).
\bibitem{ay3} S. Alexander and N. Yunes, Phys. Rev. Lett. {\bf 99}, 241101 (2007).
\bibitem{konno1} K. Konno, T. Matsuyama and S. Tanda, Phys. Rev. D {\bf 76}, 024009 (2007).
\bibitem{grumiller} D. Grumiller and N. Yunes, Phys. Rev. D {\bf 77}, 044015 (2008).
\bibitem{pretorius} N. Yunes and  F. Pretorius, Phys. Rev. D {\bf 79}, 084043 (2009).
\bibitem{konno2} K. Konno, T. Matsuyama and S. Tanda, Prog Theor. Phys. {\bf 122}, 561 (2009).
    \bibitem{olive}  B. A. Campbell, M. J. Duncan, N. Kaloper and K. A. Olive, Phys. Lett B {\bf 251}, 34 (1990).
\bibitem{reuter} M.  Reuter,  Class. Quant. Grav. {\bf 9}, 751 (1992).
\bibitem{kaloper} N. Kaloper, Phys. Rev. D {\bf 48}, 2598 (1993).
\bibitem{btz1}  M. Banados, C. Teitelboim and J. Zanelli,  Phys. Rev. Lett. {\bf 69}, 1849 (1992).
\bibitem{btz2}  M. Banados, M. Henneaux, C. Teitelboim and J Zanelli, Phys. Rev. D {\bf 48}, 1506 (1993).
\bibitem{strom}  W. Li, W. Song and A. Strominger, J. High Energy Phys. {\bf 0804}, 082 (2008).
\bibitem{mkpw} J. M. Maldacena, Adv. Theor. Math. Phys. {\bf 2}, 231
(1998); S. S. Gubser, I. R. Klebanov and A. M. Polyakov, Phys.
Lett. B {\bf 428}, 105 (1998); E. Witten, Adv. Theor. Math. Phys.
{\bf 2}, 253 (1998).
\bibitem{lemos1} J. P. S. Lemos and V. T. Zanchin, Phys. Rev. D {\bf 53}, 4684 (1996).
\bibitem{lemos2} J. P. S. Lemos, Phys. Lett. B {\bf 353}, 46 (1995).
\bibitem{lemos3} J. P. S. Lemos and V. T. Zanchin, Phys. Rev. D {\bf 54}, 3840 (1996).
\bibitem{vuorio} I. Vuorio, Phys. Lett B {\bf 163}, 91 (1985).
\bibitem{godel} K.  G\"{o}del, Rev. Mod. Phys. {\bf 21}, 447 (1949).
\bibitem{santos} C. Furtado, T. Mariz, J. R. Nascimento, A. Yu. Petrov, and A. F. Santos, Phys. Rev. D {\bf 79}, 124039 (2009).
\bibitem{ah} H. Ahmedov and  A. N. Aliev, Phys. Lett B {\bf 690}, 196	 (2010).
\bibitem{horowitz} G. T. Horowitz and D. L.  Welch, Phys. Rev. Lett. {\bf 71}, 328 (1993).
\bibitem{by}  J. D.  Brown and J. W. York, Phys. Rev. D {\bf 47}, 1407 (1993).
\bibitem{gurses} M. G\"{u}rses,  arXiv:0812.2576 [gr-qc]
\bibitem{barrow} J. D. Barrow and M. P. Dabrowski,  Phys. Rev. D {\bf 58}, 103502 (1998).
\bibitem{park} M. I. Park, Phys. Rev. D {\bf 77}, 026011 (2008).






\end{thebibliography}
\end{document}